\documentstyle[psfig,epsfig,12pt]{article}
\begin{document}
\baselineskip 0.25in
\begin{center}
\large {\bf Role of Selective Interaction in Wealth Distribution} 
\end{center}
\smallskip
\begin{center}
Abhijit Kar Gupta
\end{center}

\begin{center}
\it {\it Physics Department, Panskura Banamali College\\
Panskura R.S., East Midnapore, WB, India, Pin-721 152}\\
{\em e-mail:}~{abhijit$_{-}$kargupta@rediffmail.com}
\end{center}

\noindent {\large \bf {Abstract}}
\smallskip

\noindent In our simplified description `wealth' is money ($m$). A kinetic theory of gas like model of 
money is investigated where two agents interact (trade) selectively and exchange some amount of 
money between them so that sum of their money is unchanged and thus total money of all the 
agents remains conserved. 
The probability distributions of individual money ($P(m)$ vs. $m$) is seen to be influenced by certain 
ways of selective interactions. The distributions shift away from Boltzmann-Gibbs like 
exponential distribution and in some cases distributions emerge with power law
tails known as Pareto's law ($P(m) \propto m^{-(1+\alpha)}$). Power law is also observed in some other 
closely related conserved and discrete models. A discussion is provided with numerical support to have 
a dig into the emergence of power laws in such models.

\bigskip

\noindent{\large \bf {Introduction}}
\smallskip

\noindent 
{\it Econophysics of Wealth distributions} is an active area which involves in
interpreting and analysing real economic data of money, wealth or income distributions 
of all kinds of people pertaining to different societies and nations \cite{real}. 
A number of statistical physical models can be found in the literature \cite{eco} in 
connection with the above. Understanding the emergence of Pareto's law 
($P(m) \propto m^{-(1+\alpha)}$), now more than a century old, is one of the most important agenda. 
Some early attempts \cite{early} have been made to understand the wealth distributions, 
especially the Pareto's law where the index $\alpha$ is generally found to be in the range of 1 to 2.5 
more or less universally. 

Some recent works \cite{chak1, chak2} assume economic activities to be analogous to elastic collisions
to have Kinetic theory of gas like models proposed by Chakrabarti and group and later by 
others (We refer this class of models as `Chakrabarti model'.). 
Analogy is drawn between Money ($m$) and Energy ($E$) where temperature ($T$) is average 
money ($<m>$) of any individual at `equilibrium'. 
There has been a renewed interest in the two-agent exchange model (be it of money, energy or of 
something else) in the new scenario. 
For example, a recent work deals with social systems of complex interactions like sex which is based 
on a granular system of colliding particles (agents) with gaining energy \cite{sex}.

In this paper we deal with Chakrabarti model kind of systems where it is assumed that any two agents chosen 
randomly from a total number of agents ($N$) are allowed to interact (trade) stochastically and thus money 
is exchanged between them. 
The interaction is such that one agent wins and the other looses the same amount 
so that the sum of their money remains constant before and after interaction (trading). Therefore, it is a 
two-agent zero sum `game'. 
This way it ensures the total amount of money of all the agents ($M = \Sigma m_i$) to
remain constant. Such a model is thus a conserved model.

\bigskip

\noindent{\large \bf {The models and results}}
\smallskip

The basic steps of a money exchange (conserved) model are as follows: 

\begin{equation}
m_i(t+1)=m_i(t)+\Delta m 
\end{equation}
\begin{equation}
m_j(t+1)=m_j(t)-\Delta m,
\end{equation}

\noindent where $m_i$ and $m_j$ are money of the $i$-th and 
$j$-th agents respectively. Here we have $t$ as discrete `time' which is referred to as 
a single interaction between two agents. The amount $\Delta m$ (to be won or to be lost by an agent) is given 
by the nature of interaction. In a pure gambling, $\Delta m = \epsilon(m_i(t) + m_j(t)) - m_i(t)$, where   
stochasticity is introduced through the parameter $\epsilon$ ($0<\epsilon<1$).

If agents are allowed to interact for a long enough time, we arrive at an equilibrium 
distribution of money of individual agents. 
We arrive at a Boltzmann-Gibbs type distribution 
[$P(m) \propto {\exp(-m/<m>)}$] of individual money which is verified numerically.
This is quite the same way we arrive at the 
equilibrium energy distribution of a system of gas particles elastically colliding and 
exchanging energy with each other. The equilibrium temperature corresponds to average money, $<m>$ per agent. 

All the computer simulation results reported here, are done with system sizes 
(=total number of agents) $N=1000$. 
In all cases the system is allowed to equilibrate upto $t=10^6$ time steps. 
Averaging is done over 1000 realizations in each case. 
The final distribution of course should not depend on the initial configuration (initial 
distribution of money among the agents). The wealth distributions we deal with in this paper are 
ordinary distributions and not cumulative ones.
To obtain a distribution we take average over many different realizations; which means over a 
number of ways of the random selection of a pair of agents and also over the stochastic term $\epsilon$. 

If we intend to take average of money of a single agent over a long time, it turns out to be 
the same for all agents. 
Therefore, the distribution  
of individual time averaged money appears to be a delta function as checked by numerical
simulation. However, when the average is taken over a short time period, the delta function 
broadens to appear as a modified exponential distribution. Finally, the distribution of 
individual money at a certain time turns out to be a pure exponential one as 
mentioned earlier. This is all with randomly selected pair of agents with stochastic gain or loss (per agent). 

However, two randomly selected agents can interact with each other in various number of ways other than 
just pure gambling.
Distribution of individual money depends on how the two randomly chosen agents decide to 
interact when they meet: whether it will be random sharing of their aggregate money or 
with some rules \cite{chak1, chak2, power1}. We can in general write: 

\[\left(\begin{array}{c}
m_i(t+1) \\
m_j(t+1)
\end{array}\right)=T\left(\begin{array}{c}
m_i(t) \\
m_j(t)
\end{array}\right),\]

\noindent where $T$ is a $2\times 2$ transition matrix given by the kind of interactions. It is 
interesting to look at some properties of $T$ in connection with the emergence of distributions:
Boltzmann-Gibbs type exponential or a modified exponential 
distribution ($P(m) \propto m^{\nu}\exp(-m/<m>)$. This aspect has been investigated in our earlier work \cite{abhi}. 

Interactions or trade among agents in a society are often guided by personal choice or some social norms. 
Individual wealth distributions may be altered due to selective interactions. 
However, any arbitrary kind of selective interaction (or preferential behaviour) may have no effect. 
As an example, we examine the role played by the concept of 
family. A family in a society usually consists of more than one agent. It is quite reasonable to 
assume that the agents belonging to same family do not trade or interact among themselves. Such a  
case of selective intercation does not appear to have any influence on the 
individual wealth distribution: it remains exponentially distributed as it is checked numerically.
In this context we discuss family wealth distribution. 
In computer simulation we colour the agents belonging to a same family to keep track of. 
To find wealth distribution of families, we add up the contributions of the family members.
In {\bf Fig.1} we plot family wealth distributions 
for three cases: (i) families consist of 2 members each, (ii) 
families consist of 4 members each, and (iii) families of mixed sizes between 1 to 4. 
The distributions are clearly not pure exponential, but modified exponential distributions 
with different peaks and different widths which is quite expected (The probability of zero 
income of a family is zero.). 

\begin{figure}[htb]
\centerline{\epsfig{figure=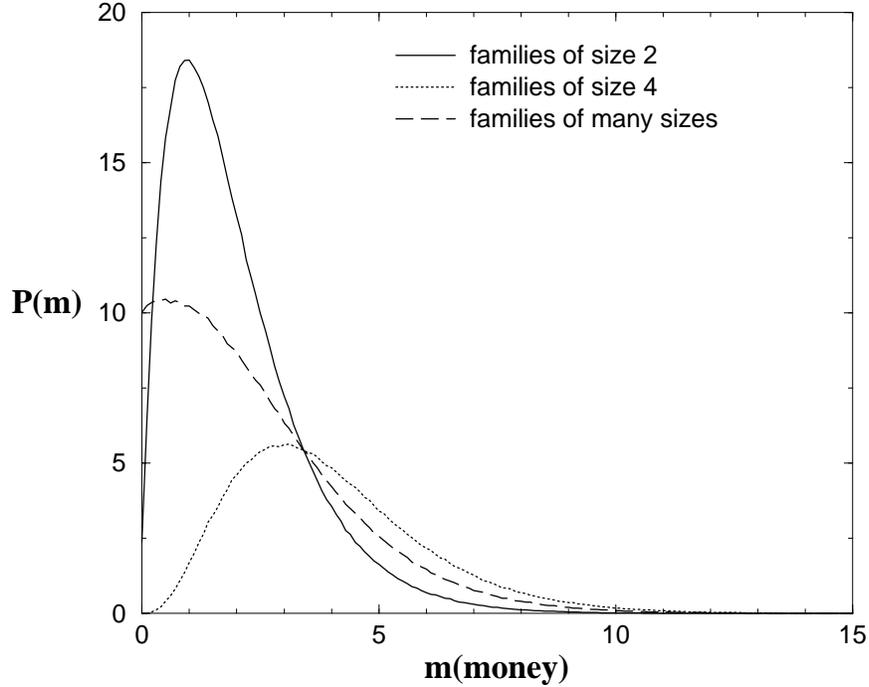, width=4.5in, height=3.7in}}
\caption{Family wealth (money) distribution: two curves are for families of all equal sizes and 
one is for families of various sizes between 1 and 4. Unit is arbitrary.}
\end{figure}

Some special ways of incorporating `selection' may play a definitive role in the individual wealth
distributions as it can be seen as follows.
Let us define a `class' of an agent by some index. The class may be understood in terms 
of their efficiency of accumulating money or some other related property. 
It is assumed that during the interactions, the agents may convert an appropriate amount of money 
proportional to their efficiency factor in their favour or against. Now the model can be understood
in terms of equations (1) and (2) with the redefined amount of exchange:

\begin{equation}
\Delta m = (\epsilon_i-1)m_i(t)+\epsilon_j m_j(t),
\end{equation}

\noindent where $\epsilon_i$'s are random numbers between 0 and 1 and are randomly assigned to the agents 
at the beginning (frozen in time).
Now let us suppose that the agents are given a choice to whom not to interact with. This option, infact, is not
unnatural in the context of a real society where individual or group opinions are important. There has been a lot 
of works on the process and dynamics of opinion formations in model social systems, a good amount of discussions 
can be found in ref.\cite{opinion}.
In our model we may imagine that the `choice' is guided by the relative class index of the two agents. We assume  
that an interaction takes place when the ratio of two class factors remain within
certain upper limit. Our requirement for interaction (trade) to happen is then
$1 < \epsilon_i/\epsilon_j <\eta$. 
Wealth distributions for various values of $\eta$ are numerically investigated. Power laws in 
the tails are obtained in all cases. 
In {\bf Fig.2} we show the distribution for $\eta = 2$. A power law is clearly observed with a power 
around 3 which means the Pareto index $\alpha$ is close to 2. 
A straight line in the log-log plot is shown for comparison (a power law with power 3.3).  

\begin{figure}[htb]
\centerline{\epsfig{figure=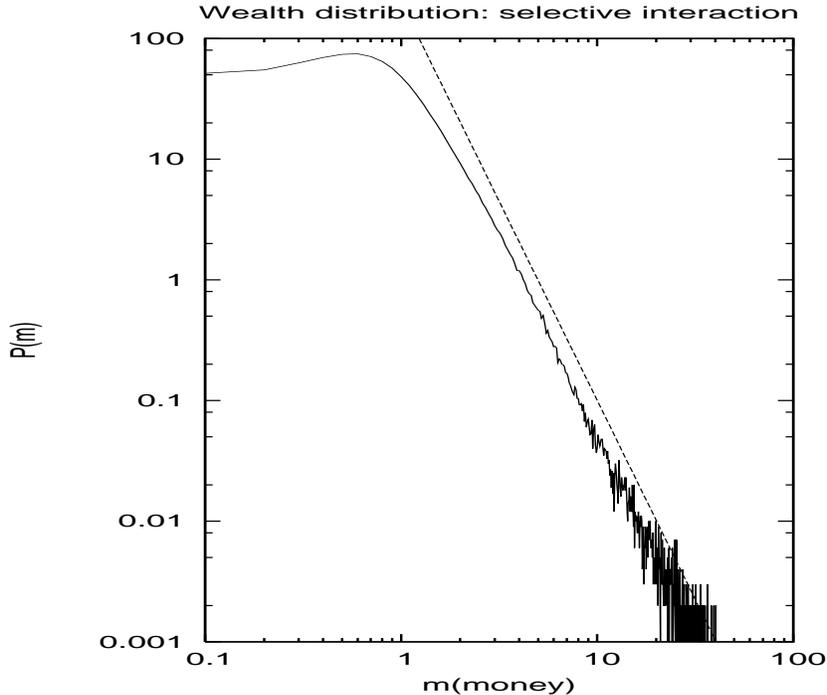, width=4.5in, height=3.7in}}
\caption{Distribution of individual money with selective interaction. 
Power law is evident in the log-log plot where a straight line is drawn with
$m^{-3.3}$ for comparison.}
\end{figure}

Many works in Econophysics or Sociophysics are devoted to understand the 
emergence of power law Pareto tails in the distrbutions \cite{power}. 
Pareto's law is observed quite universally across many nations and societies.
The power law tails in the distributions are obtained by a number of 
models like that of Chakrabarti model type (conserved) models \cite{power1} and other kinds 
of (nonconserved) models \cite{power2}. However, much is there to be understood, particularly how 
a power law emerges from simple discrete models and algorithms. 
It is known that probability distribution of money of majority is different from that of handful of 
minority (rich people). `Disparity' is more or less a reality in an economy. We may then think 
of a money exchange process within the framework of Chakrabarti model like discrete and conserved 
model in a way that the interactions among agents cause increasing variance. It is numerically examined 
whether the process of forcing the system to have ever increasing variance (measure of disparity) leads to 
a power law as power law is known to be associated with infinite variance. 
Thus we simulate a process where the random interactions (pure gambling) among 
agents [as given by equations (1) $\&$ (2)] are allowed only when the variance $\sigma = <m^2>-<m>^2$ is 
greater than the previously calculated value where the average value $<m>$ is kept fixed. 
Thus the variance attains a very high value (not shown here) with iterations under this imposed condition. 
As a result we arrive at a power law (with a low power) in the individual wealth distribution (shown in {\bf Fig.3}). 

\begin{figure}[htb]
\centerline{\epsfig{figure=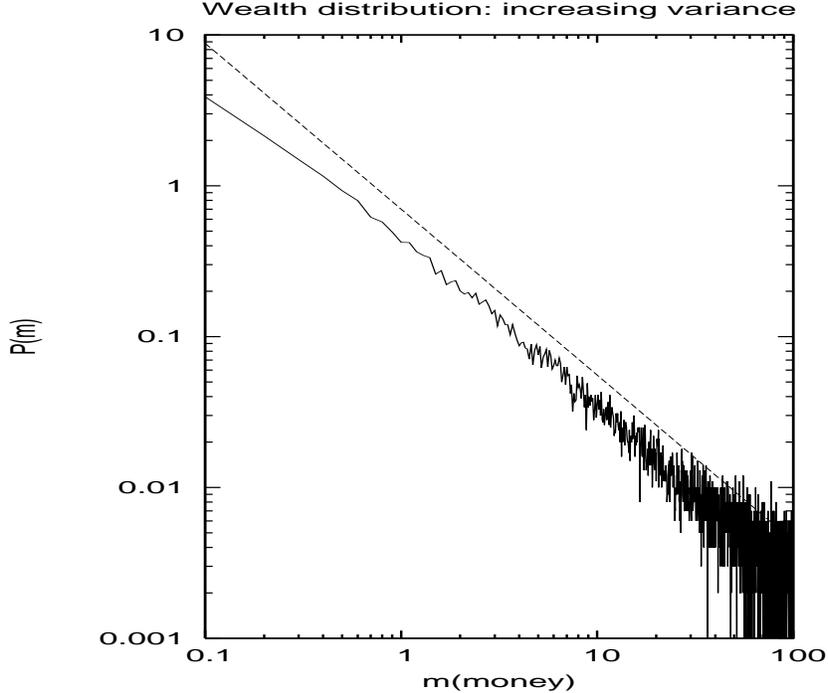, width=4.5in, height=3.7in}}
\caption{Wealth distribution with ever increasing variance. Power law is clearly seen and a 
straight line is drawn with $~m^{-1.1}$ to compare.}
\end{figure}

The above is an artificial situation though. However, we examine two models where we find that  
the money exchange algorithms are so designed that the resulting variance, in effect, increases 
monotonically with time. One such case is the model of selective
interaction as introduced before. The other prominent example is the model with random 
saving propensities proposed by Chatterjee and Chakrabarti \cite{power1},  
where $\Delta m = (1-\epsilon)(1-\lambda_i)m_i(t)+\epsilon(1-\lambda_j)m_j(t)$. 
In both cases it is checked numerically that $\sigma$ increases with time.
In {\bf Fig.4} variance ($\sigma$) is plotted with time for our model of selective interaction as 
discussed earlier. The variance $\sigma$ is seen to be monotonically increasing with time. 
In {\bf Fig.5} we show variance ($\sigma$) against time for the model of 
random saving propensity. The variance is seen to be increasing faster with time and 
attaining higher value in this case. The power law coefficients for the respective distributions may be 
related to the the magnitude of variance.
In both figures (Fig.4 and Fig.5), the scale along x-axis should be multiplied by $10^4$ to get actual time 
steps (arbitrary unit) in numerical simulation.

\begin{figure}[htb]
\centerline{\epsfig{figure=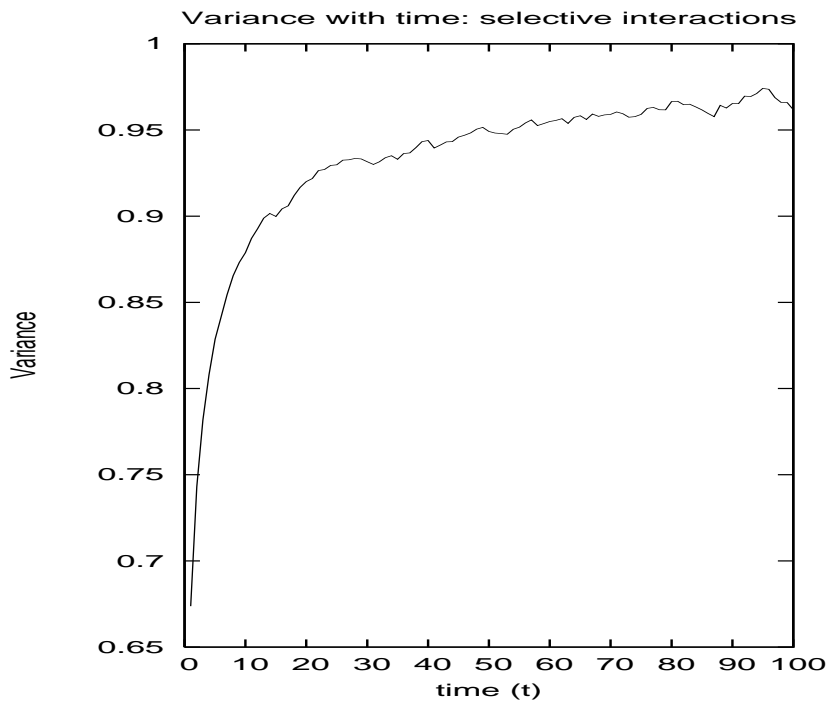, width=4.5in, height=3.7in}}
\caption{Variance against time in the model of selective interaction.}
\end{figure}

\begin{figure}[htb]
\centerline{\epsfig{figure=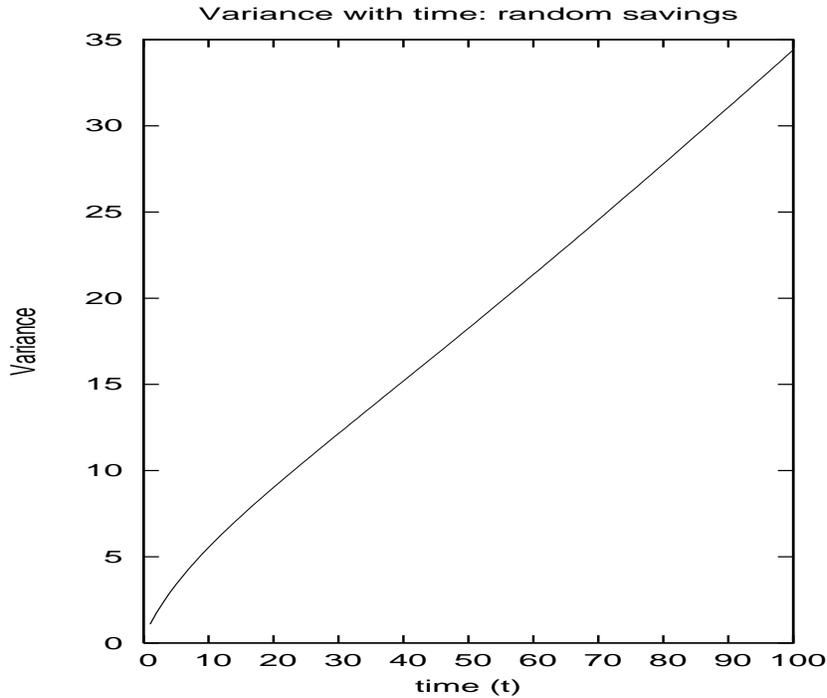, width=4.5in, height=3.7in}}
\caption{Variance against time in the model of random saving propensities}
\end{figure}

\bigskip
\noindent{\large \bf {Discussions and Conclusions}}
\smallskip

\noindent In the framework of Kinetic theory of gas like (conserved) models, random interactions 
between randomly selected pairs of agents lead to an exponential distribution of individual wealth. 
Depending on the type of interaction (that is how the money is shared among agents during an interaction), 
the distribution is altered from pure exponential to a modified exponential one and in some cases power laws 
are obtained. 
We have shown here that the distribution is also influenced by personal choice. 
Selective interaction (or preferential behaviour) of some kind (within the framework of
conserved model) can be connected to a power law in the distribution of individual wealth.
The reason for this is not very apparent though. 
However, this can be given a thought and more appropriate models may be constructed based on this information.
In a real society, people usually do not interact arbitrarily rather do so with purpose and thinking. 
Some kind of personal preference is always there which may be incorporated in some way or other.
On the other hand, a large amount of economic disparity usually exists among people. 
The detail mechanism leading to disparity is not always clear but it can be said to be associated with
the emergence of power law tails in wealth distributions. Enhanced variance is observed in our model of 
selective interaction and a power law tail is obtained in the individual wealth distribution. Monotonically 
increasing variance (with time) is also seen to be associated with another existing model (model with
random saving propensities \cite{power1}) which generates power law tails.  Therefore, there is a cause of 
thought on the inherent mechanisms in the kind of discrete and conserved models in relation to large variance 
and power law tails.  

\bigskip
\noindent {\large \bf {Acknowledgments}}

\smallskip
\noindent The author is grateful to {\it The Abdus Salam International Centre for Theoretical Physics}, 
Trieste, Italy where a major part of this work had been carried out during a short stay. 
The quality of the manuscript has been improved in a great extent 
following some suggestions and critical remarks by {\it Dietric Stauffer}. Some important references have 
also been added as he brings them into notice. The author is immensely thankful to his generosity.  
{\it Purusattam Ray} is also duly acknowledged for some interesting discussions.

\end{document}